\documentclass{article}
\usepackage{graphicx}
\usepackage{amsmath}
\usepackage{amsfonts}
\usepackage{amssymb}
\begin{document}

\title{Cloning and quantum computation}
\author{Ernesto F. Galv\~{a}o and Lucien Hardy\\ \\Centre for Quantum Computation\\Clarendon Laboratory, Department of Physics,\\University of Oxford, Parks Road, Oxford OX1 3PU U.K.}
\maketitle
\begin{abstract}
We discuss how quantum information distribution can improve the performance of
some quantum computation tasks. This distribution can be naturally implemented
with different types of quantum cloning procedures. We give two examples of
tasks for which cloning provides some enhancement in performance, and briefly
discuss possible extensions of the idea.
\end{abstract}

\section{Introduction and overview}

Since it became clear that it is impossible to make perfect copies of an
unknown quantum state \cite{Wootters Z 1982}, much effort has been put into
developing optimal cloning processes. As cloning represents a distribution of
quantum information over a larger system, it can be seen as a type of quantum
information processing tool. In this article we discuss the usefulness of
quantum cloning to enhance the performance of some quantum computation tasks.

The COPY operation in classical computing is very useful, as it allows one to
make multiple copies of the output of some computation, that can be fed as the
input to further multiple processes. In quantum computing, however, the
copying (quantum cloning) is imperfect, introducing some noise in the second
round of computation. This situation is pictured in Fig. 1(a).
\begin{figure}
\begin{center}
{\includegraphics[scale=0.4]{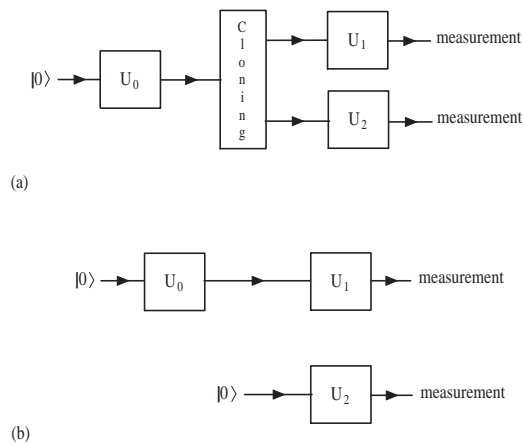}}
\caption{(a) This circuit represents the use of cloning to obtain information
about computations $U_{1}U_{0}$ and $U_{2}U_{0}$ using $U_{0}$ only once. (b)
This is one of the no-cloning strategies available for the same task. There
are other possibilities: for example, we could embed each of the $U_{j}$ in a
different quantum circuit, designed specifically for obtaining information
about $U_{j}$. }
\end{center}
\end{figure}
We can represent the first part of the quantum computation as an unitary $U_{0}$
applied to the initial state $\left|  0\right\rangle $, resulting in an output
state that we clone. We then feed the clones to two different computational
branches, represented by unitaries $U_{1}$ and $U_{2}$. At the end of the
process we make a measurement on the two final states, obtaining some
information about the two computational branches we want to perform
($U_{1}U_{0}\left|  0\right\rangle $ and $U_{2}U_{0}\left|  0\right\rangle $).
The problem with this quantum scenario is that the copies are imperfect,
resulting in lower chances of getting the correct results at the end.
Nevertheless, we will show that at least for some tasks, the use of cloning
improves our chances of correctly computing both branches, if there are
constraints on the number of times we can run the first part $U_{0}$. In the
two examples we discuss below we will be comparing approaches in which there
is some distribution of quantum information (done by the cloning process) with
any approach that does not resort to this (see Fig 1(b)).

We may discriminate between two main approaches to quantum cloning. The first
relies on adding some ancillary quantum system in a known state and unitarily
evolving the resulting combined system, deterministically obtaining a pure
state with partial mixed density matrices $\rho_{c}$ (the clones) that are as
close as possible to the original state $\left|  \psi\right\rangle $, as
measured by the fidelity $F=\left\langle \psi\right|  \rho_{c}\left|
\psi\right\rangle $ \cite{Buzek H 1996}. The clones are of the following form:%

\[
\rho_{c}=\frac{\mathbf{I}}{d}(1-\eta)+\eta\left|  \psi\right\rangle
\left\langle \psi\right|
\]
Optimal universal cloning machines are the unitaries that result in the
largest state-independent clone fidelities $F$. The efficiency of these
machines have been shown to be characterized by \cite{Werner 1998}:%

\begin{align}
\eta &  =\frac{N}{M}\frac{\left(  M+d\right)  }{\left(  N+d\right)  }%
\quad;\label{eta}\\
F  &  =\frac{(1-\eta)}{d}+\eta=\frac{M-N+N(M+d)}{M(N+d)} \label{fidelity}%
\end{align}
where $d$ is the dimensionality of $\left|  \psi\right\rangle $, $M$ is the
number of clones and $N$ is the number of original copies of $\left|
\psi\right\rangle $.

The second kind of cloning procedure is non-deterministic, consisting in
adding an ancilla, performing unitary operations and measurements, with a
postselection of the measurement results. Duan and Guo \cite{Duan G 1997, Duan
G 1998} have shown that linearly independent pure states can be
probabilistically cloned that way, and proved some theorems that allow one to
calculate the optimal efficiencies. The resulting clones are perfect, but the
procedure only succeeds with a certain probability $p<1$, which depends on the
particular set of states which we are trying to clone.

Cloning machines can be viewed as a way of encoding quantum information
contained in the input state $\left|  \psi\right\rangle $ into a number of
clones. In a sense, it accomplishes this more successfully than any procedure
that relies on obtaining information about $\left|  \psi\right\rangle $
through measurement. In order to see this, consider the universal cloning
machines described above, operating on $N$ copies of $\left|  \psi
\right\rangle $, producing $M$ identical clones described by reduced density
matrices $\rho_{c}$. Each clone has fidelity given by eq. \ref{fidelity};
notice that the fidelity of the clones is a decreasing function of $M$. The
best `classical clones' that we can produce through measurement on $\left|
\psi\right\rangle $ followed by state preparation have a lower fidelity, also
given by \ref{fidelity}, but with $M\rightarrow\infty$ \cite{Massar P 1995,
Gisin M 1997, Bruss M 1999}. It is in this sense that we can say that the
cloning process distributes quantum information about $\left|  \psi
\right\rangle $ in a way that direct measurement on $\left|  \psi\right\rangle
$ cannot.

With these considerations in mind, it is natural to wonder if, and how,
cloning can be used to improve the performance of quantum information
processing tasks. One might think that cloning could be helpful in state
estimation, but it has been shown that this task is equivalent to cloning,
when the number of copies $M\rightarrow\infty$ \cite{Bruss M 1999, Bruss E M
1998}. As a result, in order to obtain some improvement our strategy needs to
rely on using the quantum information present in the clones for further
coherent quantum information processing, in the same spirit as the circuit in
Fig. 1a. In what follows we give two examples of tasks which can be better
performed if we use quantum cloning. In the first example we apply optimal
universal cloning machines, whereas in the second we rely on the probabilistic
cloning discussed by Duan and Guo \cite{Duan G 1997, Duan G 1998}.

\section{Examples}

In this section we present two examples of quantum computational tasks whose
performance is enhanced if we distribute quantum information using quantum
cloning. The first task makes use of state-independent universal quantum
cloning, whereas the second task relies on state-dependent probabilistic
quantum cloning.

\subsection{First example}

The first example we present is based on the scenario introduced in Fig. 1a.
It models the general situation in which we want to perform $M$ different
quantum computations, all of them with some first computational steps $U_{0}$
in common. Suppose that we are constrained to run $U_{0}$ only once. This may
happen if $U_{0}$ is a complex, lengthy computation. In this case, we will be
forced to find a scheme that obtains the $M$ computation results with the
largest probability, using $U_{0}$ only once. Possible schemes may or may not
resort to cloning to distribute quantum information; the example below is one
in which cloning enables us to improve our performance, in relation to any
scheme in which there is no information distribution using cloning.

In order to specify our task, suppose that we are given $(M+1)$ quantum
blackboxes. What blackbox $j$ does is to accept one $d$-level quantum system
as an input and apply a unitary operator $U_{j}$ to it, producing the evolved
state as the output. We may think of the blackboxes as quantum oracles, or
quantum sub-computations. The $U_{j}$ are chosen randomly from all possible
$U(d)$ unitaries, using the unique uniform distribution invariant under action
of $U(d)$ (see \cite{Lubkin 1978}). Our task will be to build quantum
circuits that use each $U_{j}$ at most once to create $M$ mixed quantum states
$\rho_{j}$, each as close as possible to $\left|  \phi_{j}\right\rangle
=U_{j}U_{0}\left|  0\right\rangle ,$ $(j=1,2,...M)$, where $\left|
0\right\rangle $ is an arbitrary reference state. Our score will be given by
the average fidelity of our guesses:%

\[
\overline{F}=\frac{1}{M}\sum_{j=1}^{M}\left\langle \phi_{j}\right|  \rho
_{j}\left|  \phi_{j}\right\rangle .
\]
If we are not allowed to clone the state, there are two possible strategies.
The first no-cloning strategy is to start by finding one of the $\left|
\phi_{j}\right\rangle $, say $\left|  \phi_{1}\right\rangle =U_{1}U_{0}\left|
0\right\rangle $, with fidelity one. Now that we have used $U_{0}$ and $U_{1}$
once already, we must make guesses about the other $(M-1)$ states $\left|
\phi_{j}\right\rangle $ by using only the remaining $(M-1)$ blackboxes. As
they were drawn from an uniformly random distribution, the best we can do is
to make random guesses (each, on average, with $F=1/d$), obtaining, on
average, a score%

\[
\bar{F}_{1}=\frac{1}{M}\left(  1+\frac{(M-1)}{d}\right)  .
\]
The second no-cloning strategy starts by running $U_{0}$ , followed by
measurements that accomplish an optimal estimation of the resulting state
$U_{0}\left|  0\right\rangle $. After this, we can use the information
gathered to build the $M$ imperfect copies necessary to proceed to the second
part of the computation with the $U_{j}$ $(j=1,2,...M)$. As we have mentioned,
this second approach yields clones with fidelity given by eq. \ref{fidelity}
with $N=1,M\rightarrow\infty$ \ (see \cite{Bruss M 1999}):%

\begin{equation}
\bar{F}_{2}=\frac{2}{(d+1)}. \label{f2}%
\end{equation}
We obtain our guesses for states $\left|  \phi_{j}\right\rangle $
$(j=1,2,...M)$ by applying each $U_{j}$ $(j=1,2,...M)$ to a clone, resulting
in a score also given by eq. \ref{f2}. The best no-cloning strategy will be
either of the two presented above, depending on the parameters $M$ and $d$.

Now let us see how cloning allows us to obtain a higher score $\overline{F}$ .
We accomplish this by using a quantum circuit that first applies $U_{0}$ to
the initial state $\left|  0\right\rangle $, followed by an optimal universal
cloning machine to obtain $M$ imperfect copies $\rho_{c}$ of state
$U_{0}\left|  0\right\rangle $. We then apply each $U_{j}$ $(j=1,2,...M)$ to a
clone, obtaining reduced density matrices%

\[
\rho_{j}=\frac{\mathbf{I}}{d}(1-\eta)+\eta\left|  \phi_{j}\right\rangle
\left\langle \phi_{j}\right|
\]
with $\eta$ given by eq. \ref{eta}, with $N=1$. Using the resulting $\rho_{j}%
$'s as our guesses for states $\left|  \phi_{j}\right\rangle $ $(j=1,2,...M)$,
we obtain an overall score%

\begin{equation}
\bar{F}_{cloning}=\frac{2M+d-1}{M(d+1)} \label{fcloning}%
\end{equation}
which is always higher than $\bar{F}_{1}$ and $\bar{F}_{2}$. In fact, eq.
\ref{fcloning} represents the optimal score obtainable for this task, at least
in the case $M=2$. In order to see this, we first note that asymmetric cloning
(arising when the factors $\eta$ are in general different for each copy) is of
no help in raising the score. This can be deduced from \cite{Buzek H B 1998}
and \cite{Cerf 2000}, where the authors consider asymmetric cloning with $M=2$
and show that the sum of the fidelities of the copies is maximized by
symmetric cloning. Furthermore, the optimality of the universal cloning
procedure we have used entails optimality for the fidelity of each of the
$\rho_{j}$, and therefore a maximal value of the score $\bar{F}$. This shows
that this task is optimally performed (with optimal score given by eq.
\ref{fcloning}) if and only if we are allowed to use cloning. It is
straightforward to generalize the result to the case where we are allowed to
run $U_{0}$ $N$ times $(N<M)$, instead of just once, and quantum cloning still
offers an advantage.

The scenario described above models the situation in which we have a series of
quantum computations with some computational steps $U_{0}$ in common. We must
note that we have assumed complete lack of knowledge about the intermediate
state $U_{0}\left|  0\right\rangle $ and about the final target states
$U_{j}U_{0}\left|  0\right\rangle $ $(j=1,2,...M)$. In the general case this
will not be a good assumption, as many quantum computations will output states
picked from a limited set of states. This can be taken into account with
state-dependent quantum cloning and a different choice of scoring functions.
In the next section we give an example of this.

\subsection{Second example}

In our second example we take the blackboxes of the previous example to
consist of arbitrary quantum circuits that query a given function only once.
The query of function $f_{i}$ is the unitary that performs%

\[
\left|  x\right\rangle \left|  y\right\rangle \rightarrow\left|
x\right\rangle \left|  y\oplus f_{i}(x)\right\rangle ,
\]
where we have used the symbol $\oplus$ to represent the bitwise $XOR$
operation. For ease of analysis, we restrict ourselves to the case $M=2$ and
also restrict the set of possible functions $f_{0}$, $f_{1}$ and $f_{2}$. Our
task will involve determining two functionals, one which depends only on
$f_{0}$ and $f_{1}$, and the other on $f_{0}$ and $f_{2}$. As in the previous
example, we will compare the performances of cloning and no-cloning strategies.

In order to precisely state our task, let us start by considering all
functions $h_{i}$ which take two bits to one bit. We may represent each such
function with four bits $a,b,c$ and $d$, writing $h_{a,b,c,d}$ to represent
the function $h$ such that $h(00)=a,h(01)=b,h(10)=c,$ and $h(11)=d$. Let us
now define some sets of functions that will be helpful in stating our task:%
\begin{align}
&  S_{f0}=\{h_{0010},h_{0101},h_{1001}\},\nonumber\\
&  S_{1}=\{h_{0001},h_{0010},h_{0100},h_{1000}\},S_{2}=\{h_{0000}%
,h_{0011},h_{0101},h_{1001}\}\nonumber\\
&  S_{f12}=S_{1}\cup S_{2},\nonumber\\
&  S_{0000}=\{h_{0000},h_{1111}\},S_{0011}=\{h_{0011},h_{1100}\},
\label{2sets}\\
&  S_{0101}=\{h_{0101},h_{1010}\},S_{1001}=\{h_{1001},h_{0110}%
\},\label{2setsmore}\\
&  S_{f}=S_{0000}\cup S_{0011}\cup S_{0101}\cup S_{1001}.\nonumber
\end{align}

Now we randomly pick a function $f_{0}$ $\in S_{f0}$, after which two other
functions $f_{1}$ and $f_{2}$ are picked from the set $S_{f12}$, also in a
random fashion but obeying the constraints:%

\begin{equation}
f_{0}\oplus f_{1}\quad,\quad f_{0}\oplus f_{2}\in S_{f}\quad.
\label{constraints}%
\end{equation}
Here we use the symbol $\oplus$ to represent addition modulo 2, which is
equivalent to the bitwise $XOR$ operation. Our task will be to find in which
of the four sets $S_{0000},S_{0011},S_{0101}$ and $S_{1001}$ lie each of the
functions $f_{0}\oplus f_{1}$ and $f_{0}\oplus f_{2}$, using quantum circuits
that query $f_{0},f_{1}$ and $f_{2}$ at most once each. Our score will be
given by the average probability of successfully guessing both correctly.

The best no-cloning strategy we have found goes as follows. Firstly, note that
if $f_{0}=h_{0010}$ then both $f_{1}$ and $f_{2}$ must be in set $S_{1}$,
because of the constraints given by eq. \ref{constraints}; similarly, if
$f_{0}$ is either $h_{0101}$ or $h_{1001}$, then $f_{1}$ and $f_{2}$ must be
in set $S_{2}$. Since we have drawn the function $f_{0}$ randomly, we will
have both functions $f_{1}$ and $f_{2}$ in set $S_{2}$ with probability
$p=2/3$. We will assume that this is the case; then we can discriminate
between the two possibilities for $f_{0}$ with a single, classical function
call. Furthermore, by using the quantum circuit in Fig. 2 twice (once with
each of $f_{1}$ and $f_{2}$) we can distinguish the four possibilities for
functions $f_{1}$ and $f_{2}$.
\begin{figure}
\begin{center}
{\includegraphics[scale=0.7]{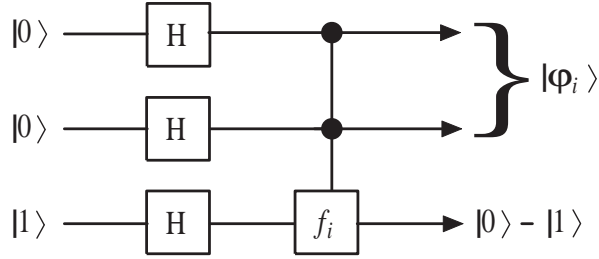}}
\caption{If function $f_{i}$ is guaranteed to be either in set $S_{1}$ or in
$S_{2}$, then this quantum circuit can be used to distinguish between the four
possibilities in each set. This is done by measuring the final state $\left|
\phi_{i}\right\rangle =\sum_{x=00}^{11}(-1)^{f_{i}(x)}\left|  x\right\rangle $
in one of two orthogonal bases, depending on which set contains $f_{i}$. The
$H$ operations are Hadamard gates.}
\end{center}
\end{figure}
This happens because this quantum circuit
results in four orthogonal states $\left|  \phi_{i}\right\rangle =\sum
_{x=00}^{11}(-1)^{f_{i}(x)}\left|  x\right\rangle $, depending on which
function in set $S_{2}$ was queried. This allows us to determine functions
$f_{0},f_{1}$ and $f_{2}$ correctly with probability $p=2/3$, in which case we
can determine which sets contain $f_{0}\oplus f_{1}$ and $f_{0}\oplus f_{2}$
and accomplish our task. Even in the case where our initial assumption about
$f_{0}$ was wrong, we may still have guessed the right sets by chance; a
simple analysis shows that our chances of getting both right this way are only
$1/16$. On average, then, by using this no-cloning strategy we obtain a score:%

\[
p_{1}=\frac{2}{3}+\frac{1}{3}\cdot\frac{1}{16}=0.6875.
\]
This is the best no-cloning score we could find for this task.

We can do better than that with quantum cloning. The idea now is to devise a
quantum circuit that queries function $f_{0}$ only once, makes two clones of
the resulting state and then queries functions $f_{1}$ and $f_{2}$, one in
each branch of the computation. Since we have some information about the state
produced by one query of $f_{0}$, the best cloning strategy will no longer be
the universal, deterministic cloning derived in \cite{Buzek H 1996} ; the
probabilistic cloning machines discussed by Duan and Guo \cite{Duan G 1997},
\cite{Duan G 1998} will suit this task better.

The quantum circuit that we apply to solve this problem is given in Fig. 3.
\begin{figure}
\begin{center}
{\includegraphics[scale=0.6]{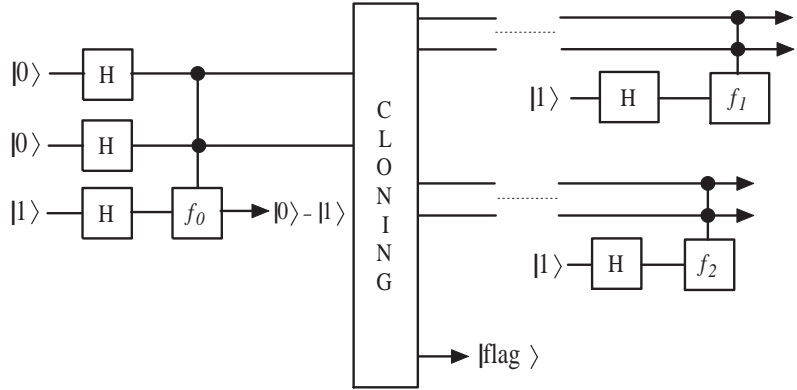}}
\caption{The cloning procedure in this circuit is probabilistic; a measurement
on the state $\left|  Flag\right\rangle $ tells us whether the cloning
succeeded. If the cloning is successful we let the clones go through the rest
of the circuit, yielding output states $\left|  \phi_{i}\right\rangle
=\frac{1}{4}\sum_{x=00}^{11}(-1)^{f_{0}(x)\oplus f_{i}(x)}\left|
x\right\rangle $, $(i=1,2)$. These states can be measured in the basis defined
by eqs. \ref{h0000}-\ref{h1001} to unambiguously decide which of the four sets
$S_{0000},S_{0011},S_{0101}$ or $S_{1001}$ contains $f_{0}\oplus f_{i}$.}
\end{center}
\end{figure}
Immediately after querying function $f_{0}$, we have one of three possible
linearly independent states (each corresponding to one of the possible $f_{0}$'s):%

\begin{align}
\left|  h_{0010}\right\rangle  &  \equiv\frac{1}{2}\left[  \left|
00\right\rangle +\left|  01\right\rangle -\left|  10\right\rangle +\left|
11\right\rangle \right]  ,\label{f0state1}\\
\left|  h_{0101}\right\rangle  &  \equiv\frac{1}{2}\left[  \left|
00\right\rangle -\left|  01\right\rangle +\left|  10\right\rangle -\left|
11\right\rangle \right]  ,\label{f0state2}\\
\left|  h_{1001}\right\rangle  &  \equiv\frac{1}{2}\left[  -\left|
00\right\rangle +\left|  01\right\rangle +\left|  10\right\rangle -\left|
11\right\rangle \right]  . \label{f0state3}%
\end{align}
We can build probabilistic cloning machines with different cloning
efficiencies (defined as the probability of cloning successfully) for each of
the states \ref{f0state1}-\ref{f0state3}. Theorem 2 of \cite{Duan G 1998}
provides us with inequalities that allow us to derive achievable efficiencies
for the probabilistic cloning process. We did a numerical search that yielded
the following achievable efficiencies for probabilistically cloning the states
in eqs. \ref{f0state1}-\ref{f0state3}:%

\begin{align}
\gamma_{1}  &  \equiv\gamma(\left|  h_{0010}\right\rangle
)=0.14165,\label{gama1}\\
\gamma_{2}  &  \equiv\gamma(\left|  h_{0101}\right\rangle )=\gamma(\left|
h_{1001}\right\rangle )=0.57122. \label{gama2}%
\end{align}
After the cloning process we can measure a `flag' subsystem and know whether
the cloning was successful or not. For this particular cloning process, the
probability of success is, on average, $p_{\text{success}}=(\gamma_{1}%
+2\gamma_{2})/3\simeq0.4280$. Let us suppose that it was successful. Then each
of the cloning branches goes through the second part of the circuit in Fig. 3,
to yield one of the four orthogonal states:%

\begin{align}
\left|  h_{0000}\right\rangle  &  \equiv\frac{1}{2}\left[  \left|
00\right\rangle +\left|  01\right\rangle +\left|  10\right\rangle +\left|
11\right\rangle \right]  ,\label{h0000}\\
\left|  h_{0011}\right\rangle  &  \equiv\frac{1}{2}\left[  \left|
00\right\rangle +\left|  01\right\rangle -\left|  10\right\rangle -\left|
11\right\rangle \right]  ,\label{h0011}\\
\left|  h_{0101}\right\rangle  &  \equiv\frac{1}{2}\left[  \left|
00\right\rangle -\left|  01\right\rangle +\left|  10\right\rangle -\left|
11\right\rangle \right]  ,\label{h0101}\\
\left|  h_{1001}\right\rangle  &  \equiv\frac{1}{2}\left[  -\left|
00\right\rangle +\left|  01\right\rangle +\left|  10\right\rangle -\left|
11\right\rangle \right]  ,\label{h1001}%
\end{align}
which can be discriminated unambiguously. We obtain state $\left|
h_{0000}\right\rangle $ if and only if the combined function $f_{0}\oplus
f_{i}$ is one of the two in set $S_{0000}$, as can be checked by calculating
the effect of the circuit in Fig. 3 for all possible $f_{0},f_{1}$ and $f_{2}%
$. The situation is similar for the other three states; the detection of each
of them signals precisely which one of the four sets $\{S_{0000}%
,S_{0011},S_{0101},S_{1001}\}$ contains $f_{0}\oplus f_{i}$ . As a result, if
the cloning process is successful, we manage to accomplish our task.

However, the cloning process will fail with probability $(1-p_{\text{success}%
})$. If this happens, a simple evaluation of the posterior probabilities for
function $f_{0}$ shows that it is more likely to be $h_{0010}$ than the other
two, thanks to the relatively low cloning efficiency for \ the state in eq.
\ref{f0state1}, in relation to the states in eqs. \ref{f0state2} and
\ref{f0state3} (see eqs. \ref{gama1}-\ref{gama2}). If we then guess that
$f_{0}=h_{0010}$, we will be right with probability%

\[
p_{0010}=\frac{(1-\gamma_{1})}{(1-\gamma_{1})+2(1-\gamma_{2})}\simeq0.5002.
\]
What is more, we are still free to design quantum circuits to obtain
information about $f_{1}$ and $f_{2}$, since at this stage we still have not
queried them. Given our guess that $f_{0}=h_{0010}$, only the four functions
in $S_{1}$ can be candidates for $f_{1}$ and $f_{2}$, because of the
constraints given by eq. \ref{constraints}. These four possibilities can be
discriminated unambiguously by running a circuit like that of Fig. 2 twice,
once with $f_{1}$ and once with $f_{2}$. The circuit produces one of four
orthogonal states, each corresponding to one of the four possibilities for
$f_{i}$. Therefore, if our guess that $f_{0}=h_{0010}$ was correct, we are
able to find the correct $f_{1}$ and $f_{2}$ and therefore accomplish our
task. In the case that $f_{0}\neq h_{0010}$ after all, we may still have
guessed the right sets by chance; a simple analysis shows that this will
happen with probability $1/16$.

The above considerations lead to an overall probability of success given by%

\[
p_{2}=p_{\text{success}}+(1-p_{\text{success}})\left[  p_{0010}+(1-p_{0010}%
)\frac{1}{16}\right]  \simeq0.7320>p_{1}=0.6875,
\]
thus showing that our cloning approach is more efficient than the previous
one, which does not use cloning. We have not proven that the first approach is
the most efficient among those that do not resort to cloning, but we
conjecture that it is.

Besides this larger probability of obtaining the correct result, our cloning
approach offers another advantage: the measurement of the `flag' state allows
us to be confident about having the correct result in a larger fraction of our
attempts. For the probabilistic cloning machines described above this fraction
was $\simeq0.428$, but \ this can be improved by choosing a different cloning
machine, characterized by $\gamma_{1}=0.3485,\gamma_{2}=0.5258$. This latter
machine signals a guaranteed correct result in a fraction $(\gamma_{1}%
+2\gamma_{2})/3\simeq0.467$ of the runs. The best no-cloning approach for
obtaining these guaranteed correct results would involve unambiguous
discrimination of the function $f_{0}$, followed by the distinction among the
four possibilities for functions $f_{1}$ and $f_{2}$ (this second step is
simple if we know $f_{0}$ for certain). Theorem 4 of \cite{Duan G 1998}
provides us with a tool to numerically determine the best efficiency for
unambiguous discrimination of $f_{0}$. A numerical search indicates that this
can be done only with efficiency $\leq1/3$, and therefore this is the limit
for the fraction of runs for which we can obtain a guaranteed correct result
for the task at hand, if we do not resort to cloning.

\section{Conclusion}

We have given two examples of tasks whose performance is enhanced by the use
of quantum cloning. As we have discussed, cloning may offer advantages for a
whole class of quantum computational tasks. Cloning need not be made only once
during the course of a computation; nor does it necessarily need be one of the
two kinds discussed above. For example, asymmetric cloning \cite{Buzek H B
1998} may also be useful, depending on the nature of the task at hand.

We must note that general quantum algorithms already manipulate quantum
information, distributing it among different parts of the quantum register
during a computation. What we have shown here is that quantum cloning can be
taken as a natural quantum information processing tool to do this quantum
information distribution, in order to optimize our use of computational
resources. It would be interesting to find other tasks that could profit from
cloning, perhaps by combining already known quantum algorithms with some
intermediate cloning steps.

In this paper we have not discussed how the cloning circuit complexity scales
with input size. Some authors have developed quantum circuits for
deterministically cloning single qubits \cite{Buzek B H B 1997, Buzek H
1998}, and networks for state-dependent cloning \cite{Chefles B 1998}. Further
work on circuits for deterministically cloning $d$-dimensional systems ($d>2$)
is still required.

\section{Acknowledgments}

We acknowledge support from the Royal Society, the ORS Award Scheme and the
Brazilian agency Coordena\c{c}\~{a}o de Aperfei\c{c}oamento de Pessoal de
N\'{i}vel Superior (CAPES).

\end{document}